# An influence of parallel electric field on the dispersion relation of graphene – a new route to Dirac logics

*Stanisław Krukowski[1, 2]\*, Jakub Sołtys[1], Jolanta Borysiuk[3,4] and Jacek Piechota[1]*

[1]Interdisciplinary Centre for Mathematical and Computational Modelling, University of Warsaw, Pawińskiego 5a, 02-106 Warsaw, Poland

[2]Institute of High Pressure Physics, Polish Academy of Sciences, Sokołowska 29/37, 01-142 Warsaw, Poland

[3]Institute of Physics, Polish Academy of Sciences, Al. Lotników 32/46, 02-668 Warsaw, Poland

[4]Faculty of Physics, University of Warsaw, Hoża 69, 00-681 Warsaw, Poland

Abstract

*Ab initio* density functional theory (DFT) simulations were used to investigate an influence of electric field, parallel to single and multilayer graphene on its electron dispersion relations close to K point. It was shown that for both single layer and AAAA stacking multilayer graphene under influence of parallel field the dispersion relations transform to nonlinear. The effect, associated with the hexagonal symmetry breaking, opens new route to high speed transistors and logical devices working in Dirac regime. The implementation of such device is presented.

*Corresponding author. Tel +48 22 8880244. E-mail address: stach@unipress.waw.pl (S. Krukowski)

1. **Introduction**

Graphene, attracting considerable interests of many researchers, became the most investigated systems in solid state physics in the recent years. This, first true two-dimensional conductor, due to its outstanding electronic properties, such as massless Dirac fermions, large electron coherence lengths, anomalous integer Hall effect[1-3, ballistic transport at the room temperature, and a good capability of integration with the silicon planar technology, is considered a key element of future high speed computing technologies, The importance of the discovery of the graphene properties was recently confirmed by 2010 Nobel Prize, awarded to its inventors A. Geim and K. Novoselov [4].

At present state of the graphene science and technology, Dirac fermions in this system present potential to be realized only. In order to exploit this potential the two basic conditions need to be fulfilled: the material should exhibit Dirac dispersion relation in the vicinity of K point in a stable way, and the device, able to open the Dirac conductive channel in fast, preferably electric way, has to be constructed. Neither of these two fundamental problems has been solved satisfactorily. Therefore successful exploitation of the graphene potential in high speed electronic devices requires removal of these two formidable obstacles. So far only in the material issue some progress has been made while in the device part no progress has been achieved.

As it is well known, single layer graphene is a perfect Dirac conductor for the case in which fermions active in the transport, have their momenta located in the vicinity of K point. In contrast to that electric properties of multilayer graphene depend on its stacking [5-7]. For both ABAB and ABCA stacking, dispersion relations are nonlinear, thus Dirac signal processing not possible using these materials. Fortunately, the AAAA stacking graphene is characterized by Dirac-type dispersion, opening possibility of Dirac logics in this system [8,9].

As freestanding single layer graphene is not mechanically stable, reliable mechanical support is needed. Such technology was developed by synthesis of large area single graphene sheets on various supports, mainly metals [10-12]. Subsequently, the graphene layer is decoupled from the support and attached onto electrically insulating material, e.g. $SiO_2$. Unfortunately, the crystallographic quality of such graphene is very low at present, therefore considerable improvement of this technology has to be attained before such route could be technically useful.

Another promising route was opened by discovery of the synthesis of graphene on a SiC substrate, by relatively simple process of Si evaporation. It was shown that graphene is created on both Si and C-faces. It was claimed that high quality graphene layers on SiC(0001), i.e. on the Si-face, smooth and homogeneouscould be synthesized, opening a possibility of fabrication of large area epitaxial graphene and its integration with the existing device technologies[12,13]. Unfortunately, in most cases four carbon ABAB stacked layers structures are created [14] thus no Dirac cones are in their band at K point [15,16]. In addition despite early claims, a strong coupling between the substrate and the single or multiple graphene layers introduces a heavy strain in the structure, inducing plethora of different defects which further deteriorates electronic properties of these layers[14-16].

Graphene could be also synthesized on C-face but the quality of the layers is evidently worse. The number of the grown layers is not limited, as the carbon bottommost layer is not attached to SiC substrate so silicon atoms could outdiffuse underneath graphene layers [17]. The new layers are nucleated randomly at whole SiC surface [18,19], therefore during coalescence they are rotated, creating turbostratic AA' stacking [20]. The electric measurements, such as Hall effect, prove that this structure is characterized by Dirac dispersion close to K-point [21,22]. It was also shown theoretically that such random arrangement of adjacent layers could results in linear dispersion typical for single layer graphene [8,9].

Alternative, technically viable new route to graphene with controlled stacking was opened when a technique of liquid exfoliation of graphene flakes was developed [23,24]. Effective method of direct exfoliation or synthesis of graphene oxide (GO) was developed [25] that could be used for production of large sheets of graphene by aggregation in various liquid solvents [26,27] unfortunately, without possibility of controlling of the obtained stacking. An important new step was made when ionic intercalation of graphene by iodine chloride (ICl) or iodine bromide (IBr) allowed to obtain large area graphene with controlled stacking [28].

In summary, several graphene synthesis techniques exists that open the possibility of manufacturing large size multilayer graphene with controlled stacking. The quality of such layers is still far from satisfactory but the route is opened for further progress in the future.

In early studies of single graphite layer (i.e. present day graphene) it was discovered that the Dirac-type dispersion relation in the vicinity of K point is intimately linked to its hexagonal symmetry [29,30]. Recent DFT results confirmed that reduction to trigonal symmetry, either by AB stacking or by hydrogen selective adsorption at every second site, leads to nonlinear dispersion relation at K point, and consequently to massive electron transport [5,9]. Such transformation cannot be exploited in high speed electronic, thus it has no technical importance. In this work we demonstrate that the other way of hexagonal symmetry breaking, by electric field parallel to graphene carbon layer, causes the same transformation. From the DFT studies of the influence of the parallel electric field it will be shown that the field opens the bandgap and transforms dispersion relation from Dirac to nonlinear in the vicinity of K point. This results is followed by the example of implementation of such transformation in electronic device that removes second major barrier on the way to high speed Dirac computer logics. Thus the last major obstacle to full use of Dirac massless transport in the computer technology is removed.

2. **Calculation method**

In this work, we report *ab initio* density functional theory (DFT) calculations to investigate the graphene-SiC system in the presence of hydrogen. In all calculations, commercially available Vienna Ab initio Simulation Package (VASP) code was employed [31-34]. The projector augmented wave (PAW) approach [35] was used in its variant available in the VASP package [34]. For the exchange-correlation functional, the local spin density approximation (LSDA) was applied. The plane wave cut-off energy was set to 500 eV. The Monkhorst-Pack mesh [36] for integration in k-space was set to 1x9x1. One or two carbon atomic layer was considered simulating single layer or bi-layer graphene. The slab replica separation empty space width was about 20 Å. At the side edges of the sheet hydrogen termination atoms are used to saturate carbon broken bonds, creating graphene ribbon.

3. **Experimental**

High-resolution transmission electron microscopy (HRTEM) observations of the graphene layers were performed using JEOL JEM 3010 transmission electron microscope operating at 300 kV. Cross-sectional TEM specimen was prepared by a standard method, based on mechanically prethinning the samples followed by an Ar ion milling procedure. The graphene layers investigated by TEM were synthesized by silicon evaporation from carbon terminated surface of *4H*-SiC (000$\bar{1}$) substrate in an Epigress VP508 SiC hot-wall chemical vapor deposition (CVD) reactor [37].

## 4. Stacking of graphene on SiC(000$\bar{1}$) surface – TEM investigations

HRTEM investigations of the majority of graphene layers synthesized on SiC substrates prove that the most abundant stacking of the graphene is AB Bernal sequence, both on silicon [14] and carbon [38] faces. The same stacking is the most stable carbon phase, also the most abundant in natural graphite. At some locations however, TEM observation of the graphene indicates on presence of hexagonal AA stacking. As presented in Fig. 1b the TEM data proves that AA stacking could be obtained during Si evaporation synthesis of thick graphene layers on C-terminated silicon carbide surface. This opens the possibility of construction of Dirac-type devices on SiC substrates.

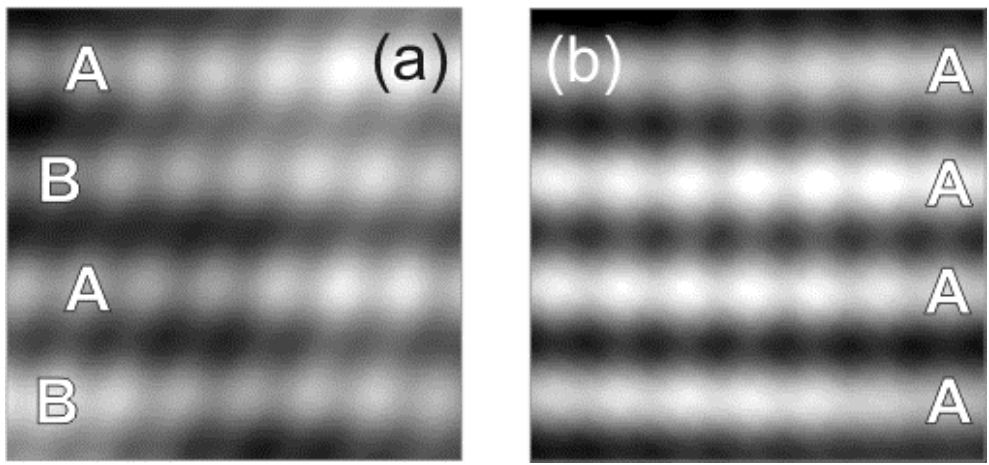

Figure 1. Cross-sectional TEM images of the graphene layers: a – AB Bernal stacking; b – AA hexagonal stacking.

The above results prove that creation of AA stacking graphene is possible at least locally by silicon evaporation from SiC(000$\underline{1}$) surface. In addition to that Hall measurement and recent DFT simulations indicate that AA' stacking, could have linear Dirac-type dispersion due to rotation of the neighboring layers. The latter result is achieved at the cost of creation of defects in the strained carbon layers which deteriorates the electric properties significantly, thus further improvement of the quality of such graphene is needed. Note that the TEM observation could not identify small rotation of the carbon layers, which may by typical for AA' stacking. Despite these shortcomings, these data indicate on the possibility of synthesis of the material having linear dispersion relations, creating potential material base for Dirac fast signal processing.

## 5. Electric field parallel to single and multiple layer graphene – DFT study

The DFT results, presented here, were obtained for the two benchmark cases: single selfstanding and two carbon layer hexagonally AA stacked graphene. The arrangement of the atoms in the slabs representing single and bi-layer graphene is presented in Fig. 2.

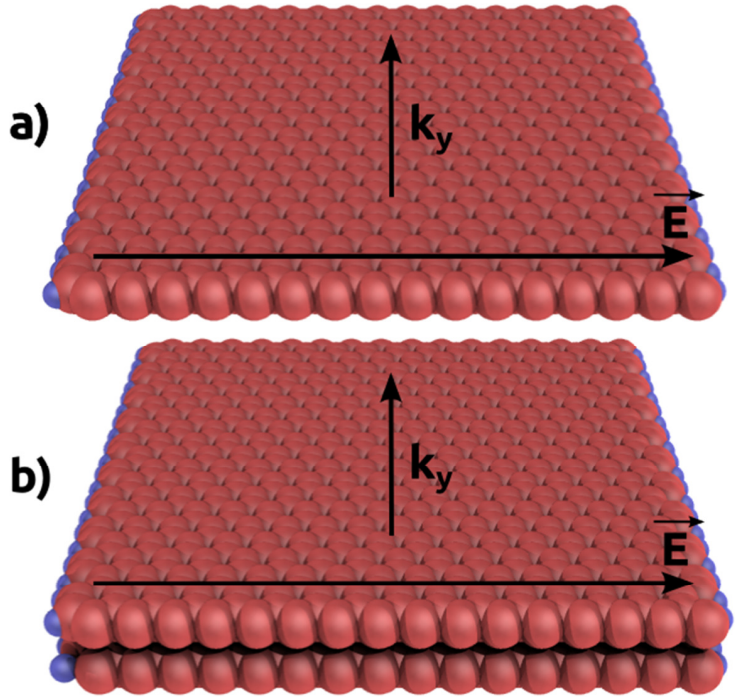

Figure 2. Arrangement of carbon atoms, denoted by red balls, in the slab used for simulation of (a) single, (b) double layer graphene in AA stacking. The broken bonds at the edges are saturated by hydrogen atoms, represented by blue balls. The slab in the field direction contains 16 unit cells of graphene, in the direction perpendicular to the field periodic boundary conditions are applied, therefore dispersion relation may be represented using wavevector denoted as $k_y$.

The carbon slab in the field direction has 16 elementary cells of graphene which, accounting the carbon-carbon interatomic distance in graphite, equal to 1.42 Å, is 39.85 Å wide. At both sides of the slab, carbon broken bonds are terminated by hydrogen atoms. In the direction perpendicular to the field the periodic boundary conditions are used. The results were obtained for no electric field and for the following parallel field intensities: 0.1 V/Å, 0.2 V/Å, 0.3 V/ Å and 0.4 V/Å, for which the total potential difference between slab edges was 3.99 V, 7.97 V, 11.96 V and 15.94 V, respectively.

The two layers slab, used in the calculation, has the interlayer separation equal to 3.39 Å which is within the typical carbon interlayer distances for graphene synthesized on silicon and carbon faces of SiC, determined by TEM to be 3.3±0.2 Å[14] and 3.4±0.2 Å [5], respectively. It is also very close to interlayer distance in graphite, equal to 3.35 Å [39]. The interlayer distance, used in the simulations, leads to large overlap of the orbitals of carbon atoms of neighboring layers that changes the dispersion relation form linear to nonlinear for AB and ABC stackings [9]. In the case of AA stacking, the above overlap, preserving linear dispersion of both branches, shifts them apart by about 0.7 eV [9]. Such carbon-carbon interlayer distance is likely in be encountered in the synthesized multilayer graphene and therefore could be used to study the transport mode in real systems.

The obtained data indicate that an increase of the electric field strength opens the bandgap at K point and transforms the dispersion to more nonlinear. In Fig. 3 the dispersion obtained for zero electric parallel field was compared with the one obtained for 0.2 V/Å case. As it was already discussed, such field intensity causes the dispersion transformation that changes massless Dirac to massive fermions. This change is associated with the opening of the gap to about 0.1 V. Such gap is sufficient to block conductance of the channel in the transistor at room temperature.

It is worth to note that the presented below dispersion diagrams are plotted along the only direction for which graphene sheet is periodic (i.e. perpendicular to the electric field). This is caused by the fact that the periodicity of the lattice in the direction parallel to the field was broken in order to prevent electric current flow in the carbon plane. Without such boundaries the field causes the electric current flow in the field direction that leads to divergence of SCF loop in DFT Kohn-Sham equation solution procedure. It fact the presented diagram is more complex than typical dispersion relation obtained for single elementary cell. Therefore the presented dispersion relation is typical for the entire graphene ribbon 16 elementary cells wide which creates overlap of many branches crowding the obtained diagrams .

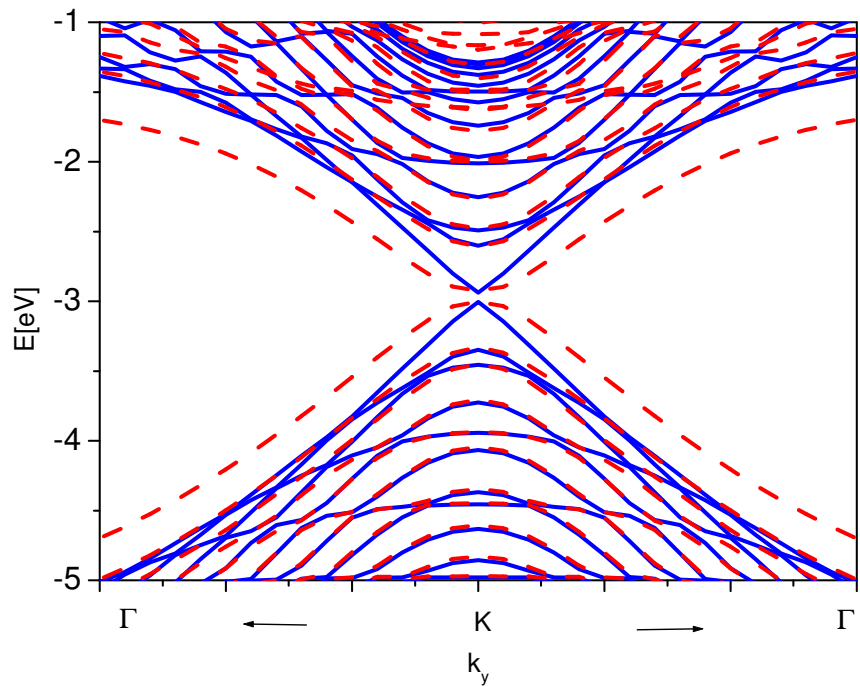

Figure 3. Electron dispersion relation for single graphene layer obtained from DFT VASP calculations for the following electric fields, parallel to the carbon plane: vanishing (blue solid line) and 0.3 V/ Å (red dashed line).

Technical implementation of the Dirac device may use multilayer AA stacked graphene, on various supports, either directly synthesized or transferred from other systems. In our simulations the influence of the support is neglected beacuse it was shown that, the graphene synthesized on C-face of SiC is coupled to the support by weak Van der Waals forces that does not change the dispersion relations [19,22,38]. In Fig. 4 dispersion relations, obtained from DFT VASP calculations for the two layers AA stacked graphene, located at the distance of 3.39 Å, are presented.

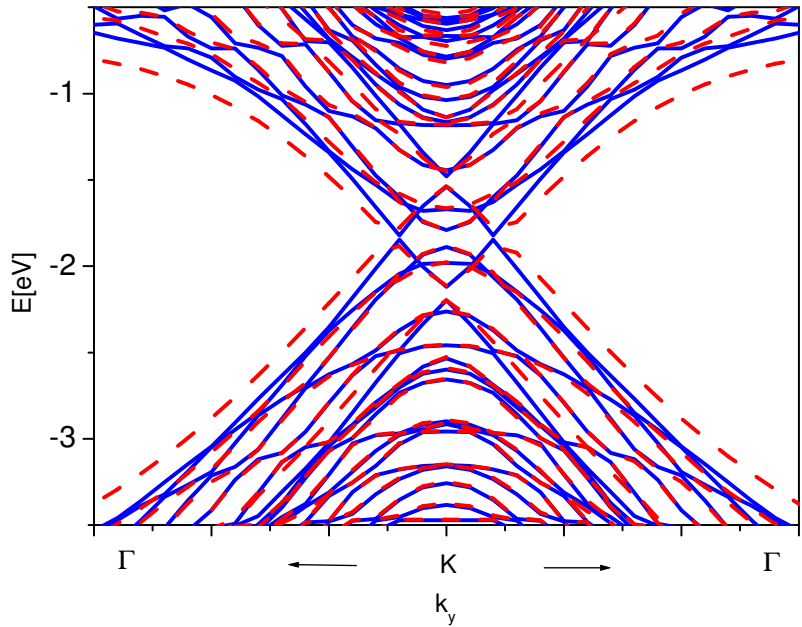

Figure 4. Electron dispersion relation for bi-layer graphene obtained from DFT VASP calculations for the following electric fields, parallel to carbon planes: vanishing (blue solid line) and 0.3 V/ Å (red dashed line).

These dispersion relations are more complicated than these above as they contain intertwined contribution from the two carbon planes. As it was shown, for zero field the two Dirac branches shifted in energy appear in the dispersion diagram [9]. Normally, these two branches are parallel and have double crossing at the Fermi level. In the present case these two branches are accompanied by the other branches that do not reach the Fermi energy level. Application of the field changes the dispersion relation again from linear to hyperbolic, in such a way creating massive fermion transport. In addition, the gap opens to the magnitude comparable to the effect observed in single layer graphene.

These data suggest that operation of the Dirac type device is possible equally the single and b-layer graphene. The basic features of the design and principle of the operation of such device is presented below.

## 6. Dirac electronic devices

The design of the fast electronic switching device has to exploit the parallel field effect that changes the dispersion from linear to nonlinear in presence of the electric parallel field. Thus the basic carbon plane should be supplemented by the two electrodes on both sides of the conductive channel. Such design is presented in Fig. 5. As the signal current is perpendicular to the field direction, this electrode need not to be electrically isolated form the carbon plane. Nevertheless, from the point of view of noise reduction it is preferable that such electrodes should be isolated.

Naturally, the presented design catches basic principle of the device only. The technical solution may include single or multiple layers graphene. On the sides of the device, the carbon plane could be cut or left continuous. The construction of the source and the drain and its connection to other electronic elements may be implemented also using different designs.

As argued before, application of the field changes the electron transport from massless to massive, bringing drastic reduction of the carrier velocity. Therefore application of the potential brings closing of the channel. In fact this may lead to slower transport only, nevertheless it allows to detect the signal or its absence in relatively short detection period. Thus the change to massive transport may lead to signal absence is short clock cycle and could be logically interpreted as channel closed.

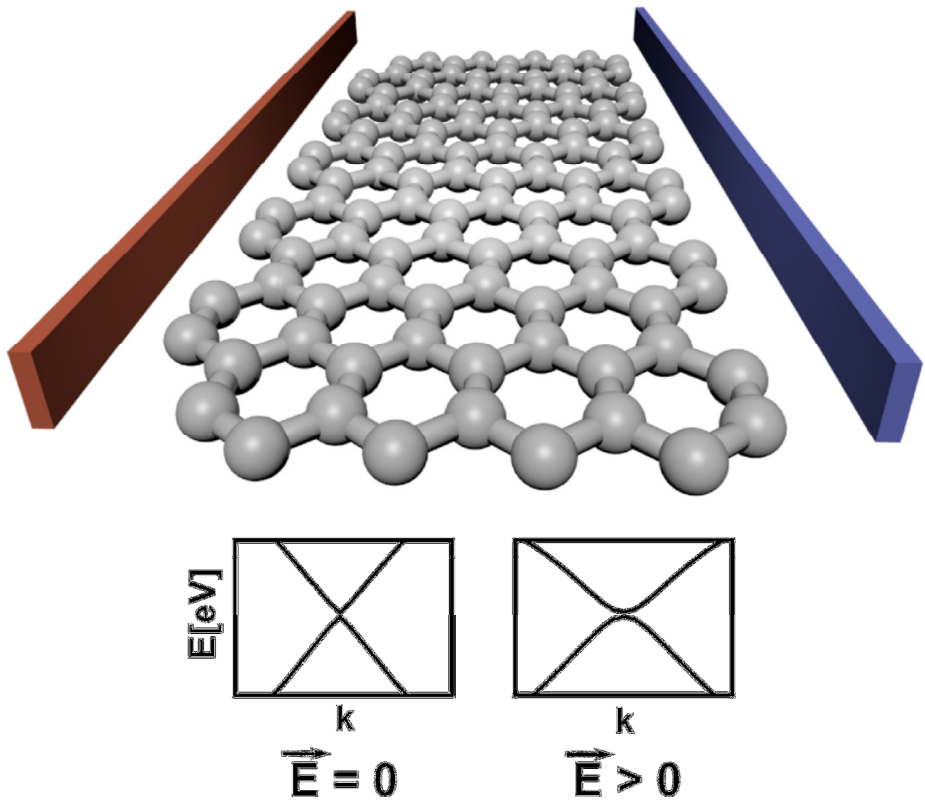

Fig. 5. Generic design of Dirac logical device, based on parallel electric field effect. Right and left: the electrodes that apply electric field to carbon plane, the carbon conduction channel is located in the middle, below – the two dispersion relations obtained for non- and vanishing electric field.

## 7. Summary

The above described principle of the design and operation of the Dirac electronic device presents viable solution that could be widely implemented in the future electronic signal processing creating the technical base for development of Dirac electronics and computer logics in particular.

DFT calculations indicated that application of the electric field, parallel to carbon plane of single or multilayer AA stacked graphene leads to transformation of the dispersion relation from linear to nonlinear in the vicinity of K point. In the case of single layer graphene this opens the bandgap, leading to complete blocking of the electric conductivity in the direction perpendicular to the field. The application of the field changes electron transport from massless to massive, dramatically slowing the motion of the carriers and transfer of the electronic signals.

The parallel field effect could be used in construction of Dirac switching devices that are electrically controlled by the two electrodes located at both sides of the channel. Such design could be implemented in the fast logical device in which the signal may be manipulated with the speed characteristic to Dirac, massless transport mode of the electrons.

**Acknowledgements**

J.B would like to thank Dr. W. Strupinski from the Institute of Electronic Materials Technology (Warsaw) for providing the graphene/SiC sample. TEM investigations were performed using the JEOL JEM 3010 microscope at the Faculty of Materials Science and Engineering of Warsaw University of Technology.

This work has been partially supported by Polish Ministry of Science and Higher Education within the SICMAT Project financed under the European Founds for Regional Development (Contract No. UDA-POIG.01.03.01-14-155/09).

**References**

[1] Berger C, Song ZM, Li XB, Wu XS, Brown N, Naud C, et al. Electronic Confinement and Coherence in Patterned Epitaxial Graphene. Science 2006; 312(5777):1191-6.

[2] Novoselov KS, Geim AK, Morozov SV, Jiang D, Katsnelson MI, Grigorieva IV, et al. Two dimensional gas of massless Dirac fermions in graphene. Nature 2005; 438(7065):197-200.

[3] Zhang Y, Tan YW, Stormer HL, Kim P. Experimental observation of the quantum Hall effect and Berry's phase in graphene, Graphene: Status and Prospects. Nature 2005; 438(7065):201-4.

[4] Geim AK. Graphene: Status and Prospects. Science 324(5934); 2009:1530-4.

[5] Borysiuk J, Soltys J, Piechota J. Stacking sequence dependence of graphene layers on SiC(000$\bar{1}$)-Experimental and theoretical investigation. J Appl Phys 2011; 109(9): 093523-1-3.

[6] Castro Neto AH, Guinea F, Peres NMR, Novoselov KS, Geim AK. The electronic properties of graphene. Rev Mod Phys 2009; 81(1):109–62.

[7] Ohta T, Bostwick A, Seyller T, Horn K, Rotenberg E. Controlling the Electronic Structure of Bilayer Graphene. Science 2006; 313(5789):951-54.

[8] dos Santos JMBL, Peres NMR, Castro AH. Graphene bilayer with a twist: electronic structure. Phys Rev Lett 2007; 99(25):256802-1-4.

[9] Soltys J, Borysiuk J, Piechota J, Krukowski S. Experimental and theoretical investigation of graphene layers on SiC(000$\bar{1}$) in different stacking arrangements J Vac Sci Technol 2012; 30(3):03D117-1-6.

[10] Suiter PW, Flege JI, Er EAS. Epitaxial graphene on ruthenium. Nature Mater 2008; 7(5):406-11.

[11] Reina A, Thiele S, Jia XT, Bhaviripudi S, Dresselhaus MS, Schaefer JA, et al. Growth of Large-area Single- and Bi-Layer Graphene by Controlled Carbon Precipitation on Polycrystalline Ni Surfaces. Nano Res 2009; 2(6):509-16.

[12] Lee S, Lee K, Zhong ZH. Wafer scale homogeneous bilayer graphene films by chemical vapor deposition. Nano Lett 2010; 10(11):4702-7.

[13] Emtsev KV, Bostwick A, Horn K, Jobst J, Kellogg GL, Ley L, et al. Towards wafer-size graphene layers by atmospheric pressure graphitization of silicon carbide. Nature Mater 2009; 8(3):203-7.

[14] Borysiuk J, Bozek R, Strupinski W, Wysmolek A, Grodecki K, Stepniewski R, et al. Transmission electron microscopy and scanning tunneling microscopy investigations of graphene on 4H-SiC(0001). J Appl Phys 2009; 105(2):023503-1-8.

[15] Mattausch A, Pankratov O. Ab Initio Study of Graphene on SiC. Phys Rev Lett 2007; 99(7):076802-1-4.

[16] Varchon F, Feng R, Hass XLJ, Nguyen BN, Naud C. Electronic Structure of Epitaxial Graphene Layers on SiC: Effect of the Substrate. Phys Rev Lett 2007; 99(12):126805-1-4.

[17] Borysiuk J, Soltys J, Bozek R, Piechota J, Krukowski S, Strupinski W, et al. Role of structure of C-terminated 4H-SiC(0001) surface in growth of graphene layers - transmission electron microscopy and density functional theory studies. Phys Rev B 2012, 85(4):045426-1-7

[18] Camara N, Huntzinger JR, Rius G, Tiberj A, Mestres N, Perez-Murano F, et al. Anisotropic growth of long isolated graphene ribbons on the C face of graphite-capped 6H-SiC. Phys Rev B 2009; 80(12):125410-1-8.

[19] Park JH, Mitchel W,C Grazulis L, Eyink K, Smith HE, Hoelscher JE. Role of extended defected SiC interface layer on the growth of epitaxial graphene on SiC. Carbon 2011; 49(2):631-5.

[20] Miller DL, Kubista KD, Rutter GM, Ruan M, de Heer WA, First PN, et al. Observing the Quantization of Zero Mass Carriers in Graphene. Science 2009; 324(5929):924-7.

[21] Plochocka P, Faugeras C, Orlita M, Sadowski ML, Martinez G, Potemski M, et al. High-energy limit of massless Dirac Fermions in multilayer graphene using magneto-optical transmission spectroscopy. Phys Rev Lett 2008; 100(8):087401-4.

[22] Orlita M, Faugeras C, Borysiuk J, Baranowski JM, Strupinski W, Sprinkle M, et al., Magneto-optics of bilayer inclusions in multilayered epitaxial graphene on the carbon face of SiC. Phys Rev B 2011 83(12):125302-1-4.

[23] Coleman JN. Liquid-phase exfoliation of nanotubes and graphene. Adv Funct Mater 2009; 19(23):3680-95.

[24] Park S, Ruoff RS. Chemical methods for the production of graphenes. Nature Nanotech 2009; 4(4):217-24.

[25] Stankovich S, Dikin DA, Piner RD, Kohlhaas KA, Kleinhammes A, Jia Y, et al. Synthesis of graphene-based nanosheets via chemical reduction of exfoliated graphite oxide. Carbon 2007; 45(7):1558-65.

[26] Hernandez Y, Nicolosi V, Lotya M, Blighe FM, Sun ZY, De S, et al. High-yield production of graphene by liquid-phase exfoliation of graphite. Nature Nanotech 2008; 3(9):563-8.

[27] Lotya M, Hernandez Y, King PJ, Smith RJ, Nicolosi V, Karlsson LS, et al. Liquid phase production of graphene by exfoliation of graphite in surfactant/water solutions. J Am Chem Soc 2009; 131(10):3611-20.

[28] Shih C-J, Vijayaraghavan A, Krishnan R, Sharma R, Han JH, Ham MH, et al. Bi- and trilayer graphene solutions. Nature Nanotech 2011; 6(7):439-45.

[29] Wallace PR. Band Theory of Graphite. Phys Rev 1947; 71(9):622-34.

[30] Slonczewski JC, Weiss PR. Band Structure of Graphite. Phys Rev 1958; 109(2):272-9.

[31] Kresse G, Hafner J. Ab initio molecular dynamics for liquid metals. Phys Rev B 1993; 47(1):558-61.

[32] Kresse G, Furthmüller J. Efficiency of ab-initio total energy calculations for metals and semiconductors using a plane-wave basis set. Comput Mat Sci 1996; 6(1):15-50.

[33] Kresse G, Furthmüller J. Efficient iterative schemes for ab-initio total energy calculations using a plane-wave basis set. Phys Rev B 1996; 54(16):11169-86.

[34] Kresse G, Joubert D. From ultrasoft pseudopotentials to t. he projector augmented-wave method. Phys Rev B 1999; 59(3):1758-75.

[35] Blöchl PE. Projector augmented-wave method. Phys Rev B 1994; 50(24):17953-79.

[36] Monkhorst HJ, Pack JD. Special points for Brillouin-zone integration. Phys Rev B 1976; 13(12):5188-92.

[37] Strupinski W, Bozek R, Borysiuk J, Kosciewicz K, Wysmolek A, Stepniewski R, et al. Growth of Graphene Layers on Silicon Carbide. Materials Res Forum 2008; 615-617:199-202.

[38] Borysiuk J, Bozek R, Grodecki K, Wysmolek A, Strupinski W, Stepniewski R, et al. Transmission electron microscopy investigations of epitaxial graphene on C-terminated 4H-SiC. J Appl Phys 2010; 108(1):013518-1-6.

[39] Delhaes P[ed] The World of Carbon: Graphite and Precursors. New York: Gordon and Breach; 2001